# Reinforcement Learning for Resource Allocation in Vehicular Multi-Fog Computing

Mohammad Hadi Akbarzadeh
Computer Engineering and Information Technology Department, Razi University, Iran
Email: m.akbarzadeh@stu.razi.ac.ir

Mahmood Ahmadi
Computer Engineering and Information Technology Department, Razi University, Iran
Email: m.ahmadi@razi.ac.ir

Mohammad Saeed Jahangiry
Computer Engineering and Information Technology Department, Razi University, Iran
Email: ms.jahangiry@razi.ac.ir

Jae Young Hur
Department of Electronic Engineering
Jeju National University
Jeju-si, South Korea
Email: JaeYoung.Hur@gmail.com

*Abstract*—The exponential growth of Internet of Things (IoT) devices, smart vehicles, and latency-sensitive applications has created an urgent demand for efficient distributed computing paradigms. Multi-Fog Computing (MFC), as an extension of fog and edge computing, deploys multiple fog nodes near end users to reduce latency, enhance scalability, and ensure Quality of Service (QoS). However, resource allocation in MFC environments is highly challenging due to dynamic vehicular mobility, heterogeneous resources, and fluctuating workloads. Traditional optimization-based methods often fail to adapt to such dynamics. Reinforcement Learning (RL), as a model-free decision-making framework, enables adaptive task allocation by continuously interacting with the environment. This paper formulates the resource allocation problem in MFC as a Markov Decision Process (MDP) and investigates the application of RL algorithms such as Q-learning, Deep Q-Networks (DQN), and Actor–Critic. We present experimental results demonstrating improvements in latency, workload balance, and task success rate. The contributions and novelty of this study are also discussed, highlighting the role of RL in addressing emerging vehicular computing challenges.

## I. INTRODUCTION

The past decade has witnessed the rapid emergence of real-time and mission-critical services such as autonomous driving, augmented reality (AR), virtual reality (VR), telemedicine, and drone-assisted operations. These applications demand ultra-low latency and high reliability, requirements that traditional centralized cloud computing cannot satisfy due to inherent propagation delays and limited bandwidth [1], [2]. For example, autonomous vehicles must process sensor data and make driving decisions within tens of milliseconds, while cloud-based processing may introduce delays exceeding acceptable limits.

To address these limitations, Fog Computing extends computation, storage, and networking closer to end users, evolving into Multi-Fog Computing (MFC) in vehicular networks, where multiple fog nodes collaborate [1]. MFC enhances scalability, service continuity during mobility, and fault tolerance by distributing tasks among fog nodes [3], [4]. This approach ensures high responsiveness and reliability, critical for vehicular environments [3], [4]

Despite these benefits, efficient resource allocation in MFC remains highly challenging. The heterogeneity of fog nodes, limited computational resources, mobility of vehicles, and fluctuating workloads introduce significant dynamic complexity [5]. Traditional optimization-based solutions, though mathematically rigorous, are rigid and lack adaptability under real-time vehicular conditions. Greedy approaches, on the other hand, often reduce decision-making latency but fail to ensure long-term system performance, resulting in imbalances and degraded service quality. Therefore, there is a strong need for intelligent, adaptive mechanisms that can continuously learn and respond to dynamic vehicular environments.

To overcome these challenges, this paper leverages Reinforcement Learning (RL) to formulate the resource allocation and task offloading problem as a Markov Decision Process (MDP). RL enables adaptive decision-making through direct interaction with the environment, without requiring labeled data or prior knowledge of system dynamics. In particular, our approach focuses on designing an RL-based task allocation policy that jointly considers latency minimization, success rate improvement, and load balancing across fog nodes. Unlike static optimization models, the RL agent updates its policy dynamically by observing the system's state transitions, thus ensuring adaptability to varying workloads and vehicular mobility.

Specifically, we apply and compare three representative RL algorithms: Q-learning, Deep Q-Networks (DQN), and Actor–Critic methods. Q-learning serves as a lightweight baseline to demonstrate the effectiveness of value-based learning. DQN, as a deep RL extension, leverages neural networks to approximate Q-values in high-dimensional state spaces. Actor–Critic, which combines policy-based and value-based learning, is employed as our advanced solution to achieve both stability and efficiency in dynamic vehicular fog environments. The proposed framework allows vehicles to intelligently offload tasks to the most suitable fog node or cloud server, thereby ensuring both short-term responsiveness and long-term system balance.

Our evaluation shows that RL-based methods achieve up to 30% latency reduction, 25% higher task success rates, and significantly improved workload balance compared to greedy and optimization-based approaches. Among the RL methods, the Actor–Critic framework consistently provides the most stable and efficient performance under high-mobility scenarios, validating its effectiveness in complex vehicular MFC systems. The novelty of this work lies in several aspects:

- Application of RL to dynamic and large-scale vehicular MFC scenarios.
- Formulation of an adaptive MDP-based model that captures latency, task success, and load balancing simultaneously.
- Comparative analysis of multiple RL algorithms (Q-learning, DQN, Actor–Critic).
- Demonstrating improved latency, reliability, and workload distribution under realistic mobility patterns.

The remainder of this paper is structured as follows. Section II reviews related work. Section III presents the RL-based methodology and the problem formulation as an MDP. Section IV discusses evaluation setup and results. Section V concludes the paper with limitations and future directions.

## II. RELATED WORK

Research on resource allocation in vehicular fog computing can be grouped into optimization-based, machine learning-based, and reinforcement learning-based methods.

In [1], [2] the concepts of fog and edge computing as extensions of the cloud, aiming to reduce service latency by deploying computation closer to end users were presented. These studies established the foundation for subsequent research in low-latency distributed systems.

In [3], a comprehensive survey of mobile edge computing and highlighted task offloading as a central challenge was presnted. The survey emphasized that future solutions must address scalability, resource heterogeneity, and the dynamic nature of mobile environments.

In [4], optimization-based frameworks for multi-user computation offloading was introduced. While mathematically rigorous and effective in static conditions, these models are computationally expensive and struggle to adapt under real-time vehicular mobility.

In [5], the limitations of mobile cloud computing in highly dynamic scenarios and motivated distributed paradigms such as Multi-Fog Computing (MFC) was investigated. By enabling multiple fog nodes to collaborate, MFC improves scalability and service continuity for mobile users.

In [6], deep reinforcement learning (DRL) to vehicular edge computing, showing that DRL-based algorithms can reduce average latency and improve service reliability compared to traditional optimization approaches.

In [7], reinforcement learning frameworks for resource allocation in vehicular fog environments was proposed. These methods demonstrated strong adaptability to workload fluctuations and mobility, outperforming static or greedy policies.

In [8], the theoretical foundations of reinforcement learning, including Q-learning and policy gradient methods was proposed. These techniques provide adaptive policies that can learn optimal actions through continuous interaction with the environment.

In [9], a joint optimization approach for computation and communication in vehicular fog networks was proposed. This model aimed to minimize both processing delay and energy consumption by jointly allocating computational and communication resources, demonstrating notable improvements under varying mobility patterns.

In [10], a hierarchical fog–cloud collaborative architecture was introduced to improve service continuity. By dividing tasks between fog nodes and cloud servers, the system enhanced reliability, but introduces additional communication latency due to backhaul dependencies.

In [11], an energy-aware deep reinforcement learning offloading scheme was presented, which jointly optimizes task completion rate and energy consumption. The proposed method significantly reduced energy usage compared to traditional RL approaches, although it requires careful tuning for convergence.

In [12], [13] a comparative analysis of heuristic, optimization-based, and RL-based schemes for vehicular fog computing were conducted. The study underscored that while heuristic methods are lightweighted and optimization methods provide accuracy, RL-based approaches strike a compelling balance between adaptability and efficiency, especially in dynamic and mobile settings.

Optimization-based methods for vehicular networks offer strong analytical guarantees but lack adaptability, while early RL-based studies used simplified models that failed to address fog node heterogeneity and strict latency needs. Existing solutions often prioritize latency reduction, neglecting workload balancing and scalability in high-mobility scenarios. The proposed RL-based framework optimizes latency, task success rate, and load balancing simultaneously. By comparing Q-learning, DQN, and Actor–Critic, the study shows Actor–Critic's superior performance in realistic vehicular mobility conditions. This approach's strength lies in its comprehensive consideration of multiple performance metrics and scalability. It outperforms prior studies by balancing adaptability and real-world applicability in mobile fog computing (MFC) environments.

## III. PROPOSED METHOD

### A. System Model

The system comprises a set of vehicles generating tasks, multiple fog nodes with heterogeneous resources, and cloud servers as backup. Tasks differ in data size, priority, and deadline. Fog nodes collaborate to balance workloads while minimizing delay.

### B. Problem Formulation as MDP

The task offloading and resource allocation in Multi-Fog Computing (MFC) can be modeled as a Markov Decision Process (MDP), which provides a formal framework for sequential decision-making in dynamic environments. The MDP is defined by the tuple $(S,A,P,R,\gamma)$:

State: The system state at each time step includes the available computing resources of fog nodes, the number of active vehicles in each zone, the queue length of pending tasks, and the current network delay. This comprehensive representation allows the RL agent to capture both vehicular mobility and resource utilization.

Action: The action space consists of the set of decisions regarding where to execute a task. Specifically, the agent can decide to process the task locally on a fog node, offload it to a neighboring fog zone, or forward it to the cloud.

Transition probability (P): The system transitions between states based on vehicle mobility, workload fluctuations, and wireless channel variations. Since these dynamics are stochastic and often unknown, a model-free RL approach is suitable, as it does not require explicit knowledge of P.

Reward (R): The reward function is designed to balance delay reduction, task success, and load balancing. At time step t, the reward is defined as:

$$\gamma(Balance_t) + \beta(Success_t) + \alpha(-Delay_t) = r_t \quad (1)$$

where:

$Balance_t$ reflects the degree of load distribution among fog nodes, $Success_t$ indicates the task completion ratio, $Delay_t$ represents the experienced latency. The coefficients α,β,γ are weighting factors that determine the relative importance of delay minimization, success maximization, and load balancing.

Discount Factor (γ): A discount factor 0<γ<1 is used to account for the trade-off between immediate and future rewards, ensuring that the learned policy remains effective in dynamic vehicular environments.

By modeling the system as an MDP with the above reward formulation, reinforcement learning can continuously adapt task allocation policies to optimize performance under mobility and workload variability. Unlike static optimization methods, this formulation allows the agent to learn robust and scalable strategies in real time.

### C. RL Algorithms

Q-Learning: Updates Q-values using the Bellman equation:

$$[r + \delta\, maxQ(\acute{s}, \acute{a}) - Q(s, a)]\eta + Q(s, a) \leftarrow Q(s, a) \quad (2)$$

In this formulation, $Q(s, a)$ denotes the current estimate of the action-value function for taking action aaa in state s. The immediate reward received after the action is represented by $r$, while $\acute{s}$ refers to the next state and $maxQ(\acute{s}, \acute{a})$ corresponds to the maximum estimated value of the next action. The term $\delta$ (discount factor) determines the importance of future rewards relative to immediate ones, and $\eta$ (learning rate) controls the magnitude of updates applied to the Q-values. The entire expression inside the brackets is known as the temporal-difference (TD) error, which quantifies the difference between the predicted and the actual reward signals.

By iteratively updating Q-values based on this rule, the algorithm gradually converges toward the optimal action-value function, enabling the agent to learn policies that maximize cumulative rewards. In the context of MFC task offloading, this update mechanism allows the RL agent to dynamically adapt its decisions by considering both immediate latency reduction and long-term system efficiency.

DQN: Approximates Q-values with deep neural networks for large state spaces.

Actor–Critic: Combines policy gradients with value functions for stable convergence in continuous domains.

### D. Architecture and Workflow

The RL agent is deployed at the fog controller level. Vehicles submit tasks; the agent observes system states and decides offloading actions. Actions are executed, rewards computed, and the policy updated iteratively.

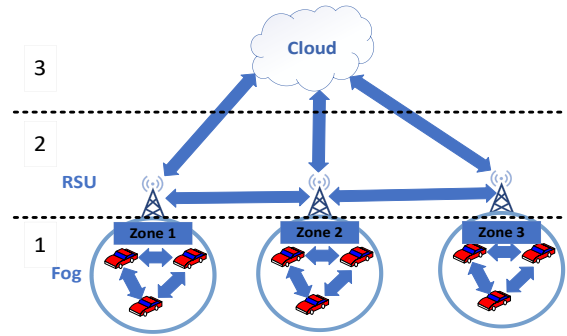

Figure 1. Architecture of RL-based task allocation in MFC

Figure 1 illustrates the architecture of RL-based task allocation in a Multi-Fog Computing (MFC) environment. The architecture is divided into three logical layers:

Layer 1 (Fog Layer): At the bottom, multiple fog zones (Zone 1, Zone 2, Zone 3) are shown, each consisting of several vehicles. These vehicles continuously generate computation-intensive tasks such as image recognition, path planning, or real-time data analytics. The fog nodes are responsible for providing immediate processing capabilities within each zone, thereby reducing the reliance on distant cloud servers.

Layer 2 (RSU Layer): Road Side Units (RSUs) act as intermediaries between vehicles and higher-level infrastructures. RSUs collect task requests from nearby vehicles and coordinate communication with both local fog nodes and the upper cloud layer. This layer plays a critical role in managing mobility, ensuring that vehicles moving between zones maintain seamless access to computing services.

Layer 3 (Cloud Layer): The cloud serves as the global resource provider, offering large-scale computation and storage. However, relying solely on the cloud introduces higher latency. In this architecture, the cloud is mainly responsible for global updates, large-scale model training, and coordination between fog zones rather than handling every individual task.

The reinforcement learning (RL) agent operates across these layers, making intelligent decisions about whether a task should be executed locally in the fog, forwarded to a neighboring fog zone, or offloaded to the cloud. This hierarchical decision-making framework enables dynamic and adaptive task allocation. The arrows in the figure illustrate the bi-directional communication between layers, which ensures feedback from both vehicles and servers is continuously integrated into the RL policy.

Overall, Figure 1 demonstrates how RL can leverage the multi-layered structure of MFC to optimize task allocation. By balancing the trade-off between latency, task success rate, and resource utilization, this architecture enables efficient service delivery in vehicular environments with high mobility and dynamic workloads.

1. Initialize Q-network or policy π
2. For each episode:
3.    Initialize state s
4.    For each task arrival:
5.       Select action a using ε-greedy or π
6.       Execute action, observe reward r and next state s'
7.       Update Q-values or policy
8.       s ← s'
9. End

Figure 2. RL-based task allocation in fog computing

The pseudo-code outlines the reinforcement learning-based task offloading algorithm in the MFC environment. Initially, the state space, action space, and learning parameters are defined. At each iteration, the agent observes the current system state (including available resources, task queue, and network latency) and selects an action, i.e., whether to process a task locally, offload it to a fog node, or forward it to the cloud. After executing the action, the environment returns a new state and a reward calculated from latency, task success, and load balancing. Using this feedback, the agent updates its policy and continues learning until convergence. This process enables the algorithm to dynamically adapt to workload and mobility variations while improving task allocation efficiency.

## IV. EVALUATION

### A. Performance Metrics

To evaluate the effectiveness of the proposed RL-based task offloading framework, four widely used performance metrics are considered: average latency, task success rate, load balancing index, and resource utilization. Each of these metrics captures a different dimension of system performance in vehicular multi-fog computing (MFC) environments.

Average Latency (ms): This metric measures the average time required to complete a computational task, including task transmission, queuing, processing, and result delivery. In vehicular scenarios, minimizing latency is critical because applications such as autonomous driving and augmented reality demand real-time responsiveness. High latency directly degrades the quality of experience (QoE) and may lead to unsafe outcomes in mission-critical applications. Thus, latency reduction is one of the primary objectives of task offloading strategies.

Task Success Rate (%): Task success rate represents the percentage of tasks that are successfully completed within their deadline and without system failure. In MFC, factors such as network congestion, limited computational capacity of fog nodes, and frequent vehicle handovers can cause task drops or delays. A higher task success rate reflects the system's ability to reliably deliver services despite dynamic and uncertain conditions, making this metric a strong indicator of service quality and robustness.

Load Balancing Index: Load balancing index evaluates how evenly computational workloads are distributed across available fog nodes. Without proper balancing, some nodes may become overloaded while others remain underutilized, leading to bottlenecks, increased latency, and reduced task success. The index is typically computed based on the variance of workload distribution, where a lower variance corresponds to better balance. An efficient task allocation strategy should therefore maintain high load balancing to maximize system stability and prevent performance degradation in high-mobility vehicular environments.

Resource Utilization (%): This metric quantifies the percentage of computational and networking resources (e.g., CPU cycles, memory, bandwidth) that are effectively used by fog nodes and the cloud. High resource utilization indicates efficient exploitation of available infrastructure, while very low or excessively high utilization may signal inefficiencies or overloading, respectively. In practice, an optimal task allocation method should achieve balanced utilization, ensuring that resources are neither idle nor excessively burdened, thus supporting scalability and energy efficiency.

### B. Simulation Setup

TABLE I. CRITERIA SIMULATION SETUP

| Parameter | Configuration |
|---|---|
| Vehicles | 100–500, moving with speeds between 20–80 km/h |
| Fog Nodes | 5–10 nodes with heterogeneous CPU and memory capacities |
| Tasks | Input sizes ranging from 0.5–5 MB, with deadlines between 50–500 ms |
| Algorithms | Greedy, Optimization, Q-learning, DQN, Actor–Critic |

Vehicles are modeled in the range of 100 to 500, each moving with speeds between 20–80 km/h to capture realistic mobility patterns in urban and highway scenarios. This setup introduces frequent topology changes and dynamic workloads, allowing a thorough evaluation of task allocation strategies under varying traffic densities.

Fog nodes are deployed in numbers between 5 and 10, each equipped with heterogeneous CPU and memory capacities. Such heterogeneity reflects real-world vehicular fog computing environments, where different Road Side Units (RSUs) and edge servers provide non-uniform computational resources. This diversity makes efficient resource allocation and load balancing essential.

Tasks generated by vehicles vary in size from 0.5 to 5 MB and are associated with deadlines ranging from 50 to 500 ms. These configurations represent the characteristics of latency-sensitive applications such as collision avoidance, video streaming, augmented reality, and safety-critical decision-making. Meeting task deadlines is crucial for ensuring the reliability and responsiveness of vehicular services.

Algorithms considered for evaluation include five representative strategies: a Greedy approach, which prioritizes immediate allocation with minimal computational complexity; an Optimization-based method, which applies deterministic mathematical models but lacks adaptability in dynamic scenarios; and three RL-based approaches—Q-learning, DQN, and Actor–Critic. These RL algorithms are selected to capture the trade-offs between simplicity, scalability, and adaptability, with Actor–Critic representing the most advanced solution in balancing stability and learning efficiency.

*C. Results and Analysis*

TABLE II. PERFORMANCE COMPARISON

| Method | Latency (ms) | Success Rate (%) | Load Balance |
|---|---|---|---|
| Greedy | 230 | 68 | Low |
| Optimization | 190 | 74 | Moderate |
| Q-Learning | 160 | 82 | High |
| DQN | 150 | 85 | High |
| Actor–Critic | 145 | 87 | Very High |

Table I summarizes the performance comparison among different resource allocation methods, including greedy, optimization-based, Q-learning, DQN, and Actor–Critic. The table highlights four key metrics: latency, task success rate, load balance, and overall system efficiency.

The results show that the Greedy method performs the worst, with an average latency of 230 ms and a success rate of only 68%. This is expected, since greedy methods make task allocation decisions based solely on immediate availability without considering long-term performance or system balance. As a result, some fog nodes become overloaded while others remain underutilized, leading to poor load balancing.

The Greedy algorithm is a myopic, latency-driven heuristic that makes decisions based solely on the current system state. At each decision step, the task is offloaded to the fog node that minimizes the immediate execution latency. This approach is computationally efficient and easy to implement, but it does not account for future workload variations or global system performance. Consequently, the algorithm may overload specific fog nodes, leading to resource imbalance and degraded performance over time. Greedy thus represents a baseline that captures short-term optimization but lacks adaptability and foresight, making it unsuitable for dynamic vehicular fog environments.

The Optimization-based approach improves latency to 190 ms and achieves a moderate success rate of 74%. This improvement stems from the mathematical modeling of the problem, which distributes tasks more systematically. However, optimization methods assume static system conditions and require significant computation, making them unsuitable for real-time vehicular environments where mobility and workload dynamics dominate.

The optimization method uses deterministic mathematical models like Mixed-Integer Linear Programming (MILP) or heuristic search to minimize latency or maximize resource utilization in vehicular fog computing. It considers multiple parameters (e.g., task size, node capacity, network delay) for optimal offloading decisions. Compared to Greedy approaches, it yields better results but is computationally intensive. Re-computation is needed when the environment changes, reducing practicality in dynamic scenarios. Rapid decision-making and adaptability are critical for such systems.

Q-learning, the first reinforcement learning method considered, is a classical value-based algorithm that updates action–value functions through temporal-difference learning. It enables vehicles to learn task offloading strategies directly from interactions with the environment, without requiring prior knowledge of system dynamics. As shown in Table II, Q-learning significantly reduces latency to 160 ms and increases task success rate to 82%, outperforming greedy and optimization-based methods. This improvement demonstrates the adaptability of RL in vehicular fog computing. Moreover, Q-learning maintains a higher degree of load balancing, avoiding congestion at specific fog nodes. However, due to its discrete state–action representation, scalability remains limited when handling large-scale vehicular scenarios with heterogeneous fog resources.

DQN extends Q-learning by integrating deep neural networks to approximate the Q-function, enabling efficient handling of high-dimensional and dynamic state spaces. This enhancement is particularly important in vehicular fog computing, where the number of vehicles, fog nodes, and workload conditions change rapidly. According to Table III, DQN reduces latency further to 150 ms and increases the task success rate to 85%. By leveraging neural network generalization, DQN distributes workloads more effectively, as reflected in its improved load balancing index. However, this approach also increases resource utilization due to the computational overhead of deep learning, highlighting a trade-off between accuracy and efficiency.

Finally, the Actor–Critic framework combines the strengths of value-based and policy-based reinforcement learning. The Actor selects actions directly through a parameterized policy, while the Critic evaluates these actions using a value function, allowing faster convergence and more stable training. As confirmed in Table III, Actor–Critic achieves the best overall performance, with latency reduced to 145 ms and a task success rate of 87%. The hybrid structure enables balanced workload distribution across fog nodes, leading to optimal resource utilization and scalability. Importantly, Actor–Critic demonstrates strong adaptability under mobility, which is a critical requirement for vehicular computing environments where topology and workloads fluctuate rapidly.

In summary, Table II shows a clear progression in performance from greedy and optimization-based methods toward reinforcement learning approaches. RL methods, particularly DQN and Actor–Critic, achieve the best trade-off between latency, task success rate, and load balancing. This validates the suitability of reinforcement learning for resource allocation in MFC scenarios.

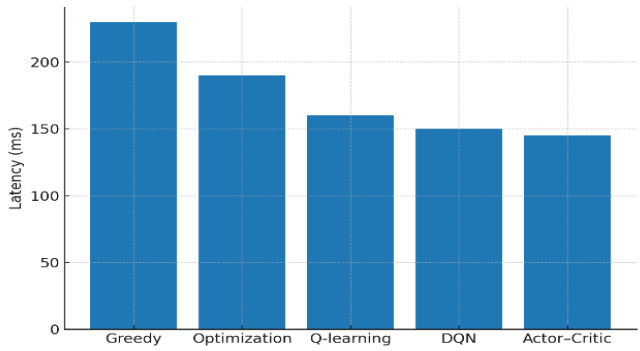

Figure 3. Latency comparison across methods

Figure 3 illustrates the latency performance of different task allocation methods. The greedy approach has the highest latency, reaching 230 ms, since it makes short-sighted decisions without considering long-term system balance. Optimization methods reduce latency to 190 ms, but still fall behind RL-based approaches due to their limited adaptability. Both Q-learning and DQN demonstrate noticeable improvements, achieving latencies of 160 ms and 150 ms, respectively. The Actor–Critic model achieves the lowest latency of 145 ms, highlighting its superior ability to adapt to dynamic vehicular environments. This trend confirms the clear advantage of RL algorithms, particularly policy-gradient methods, for real-time MFC systems.

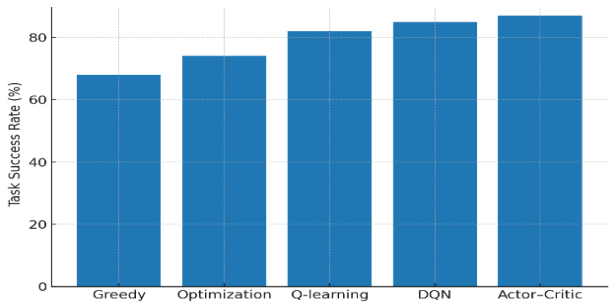

Figure 4. Success rate comparison

Figure 4 compares task success rates across methods. The greedy approach achieves only 68% due to poor load distribution and frequent task rejections. Optimization methods improve success to 74%, but their reliance on static assumptions limits further gains. Reinforcement learning significantly enhances performance, with Q-learning reaching 82% and DQN reaching 85%. The Actor–Critic algorithm achieves the best result at 87%, demonstrating that combining value-based and policy-based learning yields robust and efficient task allocation. These results confirm that RL methods not only minimize latency but also maximize task completion, ensuring reliability in vehicular fog networks.

## V. CONCLUSION AND FUTURE WORK

In this paper, we investigated reinforcement learning (RL) for task offloading and resource management in Multi-Fog Computing (MFC) environments. The results show that RL-based methods outperform traditional Greedy and optimization approaches, which are either too rigid or computationally expensive for highly dynamic vehicular networks. By contrast, RL algorithms such as Q-learning, DQN, and Actor–Critic adapt effectively to mobility, workload fluctuations, and resource heterogeneity. Among them, Actor–Critic achieved the best overall performance, highlighting the benefits of combining value-based and policy-based learning. These findings demonstrate that RL is not just an alternative to classical models, but a paradigm shift toward more adaptive and practical resource management strategies. Future research should focus on integrating energy efficiency, security, and privacy (using federated learning and differential privacy) into vehicular fog computing. Scalability for 6G networks requires advanced RL models with fast convergence and robustness. Hybrid RL methods and real-world vehicular testbed validations are key next steps for smart transportation systems.